\documentclass[journal=jpclcd,manuscript=article, layout=twocolumn]{achemso}

\usepackage[version=3]{mhchem} 
\usepackage{xcolor} 
\usepackage{multirow}
\usepackage{amsmath}
\usepackage{amsfonts}
\usepackage{pdfpages} 
\usepackage{afterpage}

\newcommand\blankpage{%
    \null
    \thispagestyle{empty}%
    \addtocounter{page}{-1}%
    \newpage}

\author{Amrita Goswami}
\affiliation[University of Iceland]
{Science Institute and Faculty of Physical Sciences, University of Iceland, VR-III, 107 Reykjav\'{\i}k, Iceland}

\author{Alejandro Pe\~{n}a-Torres}
\affiliation[University Iceland]
{Science Institute and Faculty of Physical Sciences, University of Iceland, VR-III, 107 Reykjav\'{\i}k, Iceland}

\author{Elvar \"O. J\'{o}nsson}
\affiliation[University of Iceland]
{Science Institute and Faculty of Physical Sciences, University of Iceland, VR-III, 107 Reykjav\'{\i}k, Iceland}

\author{Sergei A. Egorov}
\affiliation[University Iceland]
{Science Institute and Faculty of Physical Sciences, University of Iceland, VR-III, 107 Reykjav\'{\i}k, Iceland}
\alsoaffiliation[University of Virginia]
{Department of Chemistry, University of Virginia, Charlottesville, Virginia 22901, USA}

\author{Hannes J\'{o}nsson}
\affiliation[University of Iceland]
{Science Institute and Faculty of Physical Sciences, University of Iceland, VR-III, 107 Reykjav\'{\i}k, Iceland}
\email{hj@hi.is}

\title[An \textsf{achemso} demo]
{
Evidence of sharp transitions between octahedral and capped trigonal prism states of the solvation shell of the Fe$^{+3}$(aq) ion
}

\abbreviations{MD, DFT, IOD}

\begin{document}

\begin{tocentry}

\includegraphics[width = \linewidth]{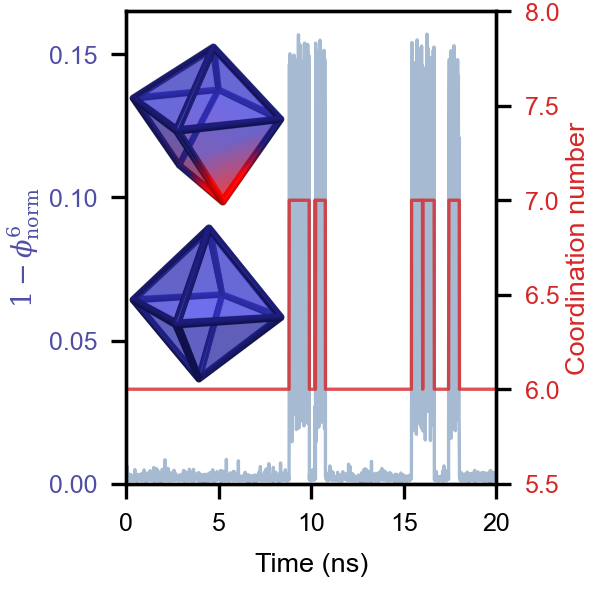}

\end{tocentry}

\begin{abstract}
The structure of the solvation shell of aqueous Fe$^{+3}$ ion has been a subject of controversy due to discrepancies between experiments and different levels of theory. We address this issue by performing simulations for a wide range of ion concentrations, using various empirical potential energy functions, as well as  density functional theory calculations of selected configurations. The solvation shell undergoes abrupt transitions between two states: an octahedral (OH) state with 6-fold coordination, and a capped trigonal prism (CTP) state with 7-fold coordination.
The lifetime of these states is concentration dependent. In dilute $\mathrm{FeCl_3}$ solutions, the lifetime of the two states is similar ($\approx 1$ ns). However, the lifetime of the OH state increases with ion concentration, while that of the CTP state decreases slightly. When a uniform negative background charge is used instead of explicit counterions, the lifetime of the OH state is greatly overestimated.
These findings underscore the need for further experimental measurements as well as high-level simulations over sufficiently long timescales and low concentration.
\end{abstract}

One of the fundamental tasks in physical chemistry is to gain an understanding of ion solvation, especially in water solutions.  A key issue is the solvent shell that forms around the ions, which can dictate their chemical and physical properties \cite{Ohtaki1993}. Experimental measurements of solvation shells are challenging, and computer simulations can provide complementary microscopic insight into their structure and dynamics. 
%
Over the past two decades, several potential energy functions have been developed to model ion-water interactions \cite{Amira2004,qing2013local,Li2013a, Chillemi2002, Joung2009, Duarte2014}. Simulations based on recently developed functions for describing the interaction between ions and rigid point charge water models have been able to successfully reproduce both the experimentally measured solvation free energy and the ion-oxygen distance for a wide range of aqueous ionic solutions \cite{Li2021a, Zhang2021, Li2020a, Zhang2021a, Li2014, Li2013}. 
While solvation shells of aqueous transition metal ions typically correspond to sixfold coordination in an octahedral structure, 
sevenfold coordination shells have, in some rare cases, also been reported \cite{Lewis1975, Yi2020, Haiges2007, Lin1996,Huang2010, migliorati2013quantum}. 
For example, both theoretical and experimental evidence has been presented for the existence of a flexible hepta-coordinated Hg$^{+2}$(aq) ion \cite{Chillemi2007}. 

A particularly relevant system is the solvation of Fe$^{+3}$, for which the prevailing view 
of the solvation shell is a well-defined octahedral arrangement of six $\mathrm{H_2O}$ molecules \cite{Richens1997}.
%
%
However, simulations based on empirical potential functions yield larger coordination numbers, ranging from 6.3 to 6.9.
This has been dismissed as an 
artefact of 
using 
point charges and the 12-6 or 12-6-4 functional form.\cite{Li2021a, Zhang2021, Li2014} 
 
As a result, there have been various attempts to explicitly restrain the coordination number to 6.0. 
For instance, the atomic charges on the six neighbouring water molecules have been modified to obtain a hexa-coordinated solvation shell, but this has the unfortunate side effect of eliminating the possibility of ligand exchange \cite{Li2021}. 
Alternatively, the positive charge of the Fe$^{+3}$ ion has been distributed on six sites near the central metal ion in the desired predefined octahedral coordination geometry \cite{Aaqvist1990, Liao2017}. This class of models allows for the possibility of ligand exchange
but has the drawback that
different types of solvation shells, such as the experimentally observed tetrahedral $[\mathrm{FeCl_4}]^-$ ion, cannot be reproduced \cite{Magini1979, Magini1982, Giubileo1983, Gammons2022}. 

Experimental measurements on ferric chloride solutions  
have led to 
a plethora of reported solvation structures, 
including the tetrahedral $[\mathrm{FeCl_4}]^-$, octahedral \emph{trans}-$[\mathrm{FeCl_2(H_2O)_4}]^+$, $[\mathrm{FeCl_3(H_2O)_3}]$, and $[\mathrm{FeCl(H_2O)_5}]^{2+}$ \cite{Brady1964, Lind1967, Wertz1980, Wertz1981, Luter1981, Magini1979, Magini1982, Giubileo1983, Brady1958, Standley1961, Gammons2022}. 
The experimental results indicate that the solvation shell structure is strongly dependent on ion concentration\cite{Persson2018, Smirnov2019, Baumler2019}. A point of some contention is the identity of the dominant $\mathrm{Fe^{3+}}$ form in concentrated ferric chloride solutions, with ion concentration on the order of 1 $\mathrm{mol/dm^3}$. A recent EXAFS (Extended X-Ray Absorption Fine Structure) 
concludes that the prevailing species is \emph{trans}-$[\mathrm{FeCl_2(H_2O)_4}]^+$ at this high concentration, while for a low concentration of  0.001 $\mathrm{mol/dm^3}$ the measurements are interpreted to be consistent with $[\mathrm{Fe(H_2O)_6}]^{3+}$.\cite{Persson2018}  
X-ray and neutron diffraction experiments on dilute solutions can be impeded by low signal intensity; however, EXAFS is considered to be a suitable technique to examine the structure of dilute aqueous ionic solutions \cite{munoz1995exafs}. While 
this
and other measurements of low concentration solutions\cite{thompson2001x, Ohtaki1993} point to the conclusion that the solvation shell consists only of H$_2$O, it is not clear whether the coordination number is precisely 6.0 \cite{Ohtaki1993}.  
%


Unlike the calculations based on potential energy functions, the simulations where energy and atomic forces are obtained from 
density functional theory (DFT), 
using either the PBE\cite{sit2006realistic,Bogatko2010} or B3LYP\cite{Lei2012,Nazmutdinov2019,Mandal2022}
functional approximations, have not shown deviations from the octahedral form of the solvation shell.
\cite{Lei2012,sit2006realistic,Nazmutdinov2019,Mandal2022,Bogatko2010}
As the computational effort is large, the time interval represented by such simulations has necessarily been short, spanning at most a few tens of ps, and an indication of a change in the structure of the first solvation shell has not been found.\cite{Mandal2022,Bogatko2010}
The simulations have, furthermore, been carried out using a uniform negative background charge instead of explicit counterions and for a high $\mathrm{Fe}^{3+}$ concentration corresponding to a $[\mathrm{Fe}^{3+}]:[\mathrm{H_2O}]$ ratio ranging from 1:18 to 1:137. 
These are much higher ion concentrations than have been used in the simulations based on potential energy functions, 1:600\cite{Zhang2021} to 1:1085.\cite{Li2021a, Li2014}

In the present study,
we address the disparity between DFT calculations and simulations with potential energy functions, cognizant of differences in ion concentration and simulation conditions. We systematically investigate the structure of the solvation shell of $\mathrm{Fe}^{3+}$(aq), using a sphericity measure. The simulations are performed 
over long time intervals, with several potential functions, for a wide range 
in ion concentration.
We also compare the effects of using explicit counterions (the recommended practice) with the use of a uniform negative background charge 
(used in the higher-level simulations, so far).

The results show remarkably clear and abrupt changes between two states of the solvation shell: an octahedral (OH) state, corresponding to sixfold coordination, and a capped trigonal prism (CTP) state with 7-fold coordination.
The existence of these two distinct states challenges the conventional premise of a single octahedral state. 
The effects of ion concentration and the presence of explicit counterions versus uniform background charge on the solvation structure are 
found to be strong. 
This could possibly help
reconcile the reported controversy between different types of simulation approaches.
We also obtain energy-minimized CTP and OH structures from DFT calculations of selected configurations, in addition to the energy barrier between them.  



In this work, we present a new and robust order parameter for distinguishing between OH and non-octahedral structures, which we use in the analysis of the simulations. The structure of the solvation shell 
is
characterized by a sphericity-based order parameter.
The sphericity quantifies the
similarity of a 3-dimensional shape to a perfect sphere. It is defined as the ratio of the surface area of a sphere with the same volume as the shape considered, to the actual surface area of the shape\cite{Wadell1935}

\begin{equation}\label{eqn:sphericity}
\phi = \frac{\pi^{\frac{1}{3}} (6 V_s)^{\frac{2}{3}} }{A_s}, \tag{$1$}
\end{equation} 
where $V_s$ and $A_s$ are the volume and surface area of the shape, respectively. 

Originally formulated for the analysis of sedimentary rock fragments \cite{Wadell1935}, the sphericity has been used in diverse applications, including nanoparticle synthesis \cite{Sau2001,Stevenson2012,Mo2022}, nanofluid thermal conductivity \cite{You2022,Nimmagadda2016}, nanoparticle aggregation in fluidized beds \cite{Hakim2005}, hypersensitivity reactions to liposomal drugs \cite{Szebeni2012,Stater2021}, cancer research \cite{Li2020,Davey2020,Szeto2009,Kingston2019,Shin2021}, and nanomedicine \cite{Baysal2016}. 

We define $\phi^{6}$ as the sphericity of the convex hull formed by the closest six neighbouring oxygen atoms of the Fe ion center. The sphericity of a perfect octahedron is $0.846$ \cite{Li2012,Bullard2013,Zheng2015} and we divide by this number to define a renormalized sphericity, $\phi^{6}_{\mathrm{norm}}$. For a polyhedron with $6$ vertices, $\phi^{6}_{\mathrm{norm}}$ can therefore have a maximum value of $1.0$. 

\begin{figure}
  \centering
  \includegraphics[width = \linewidth]{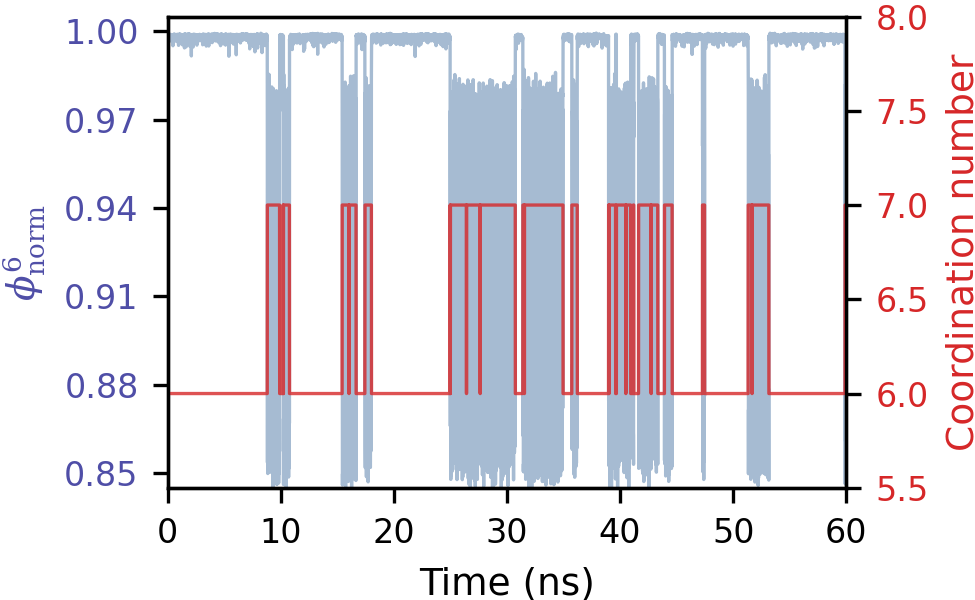}
  \caption{
Time evolution of the coordination number (red) and the normalized sphericity (blue), $\phi^{6}_{\mathrm{norm}}$, of the convex hull formed by the first six neighbours of the $\mathrm{Fe}^{3+}$(aq) ion in a room temperature simulation of a FeCl$_3$ solution with concentration corresponding to a $[\mathrm{Fe}^{3+}]:[\mathrm{H_2O}]$ ratio of 1:597.
Both measures of the solvation shell structure show abrupt transitions between sustained states
corresponding to
octahedral and capped trigonal prism configurations. The average coordination turns out to be 6.4 in this case.   
}
  \label{fig:figCoord}
\end{figure}

\begin{figure*}
  \centering
  \includegraphics[width = \linewidth]{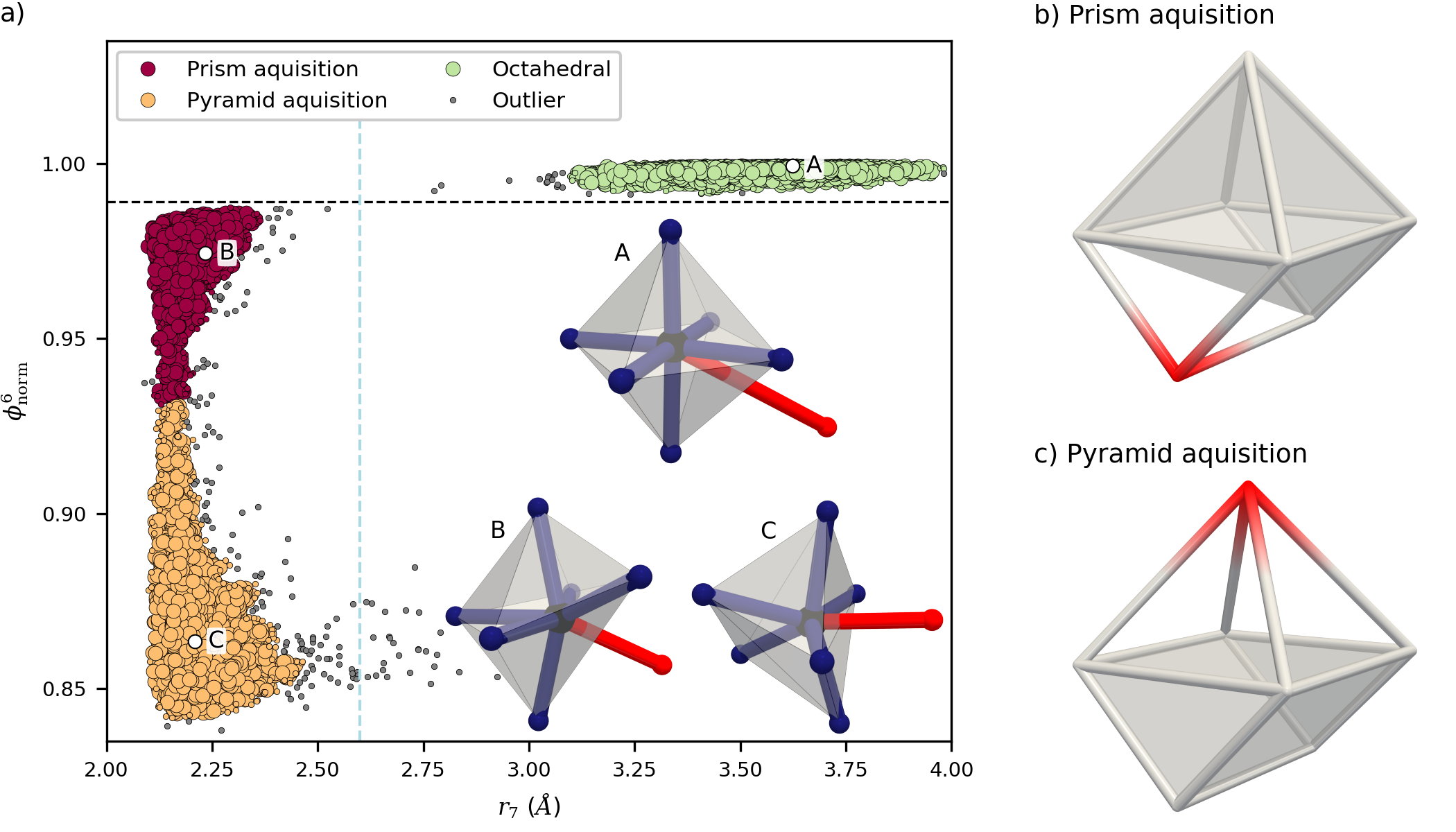}
  \caption{
(a)  The normalized sphericity, $\phi^{6}_{\mathrm{norm}}$, as a function of the distance to the seventh water molecule from the $[\mathrm{Fe}^{3+}]$ ion, $r_{7}$. Each point is coloured according to its inclusion in a particular 
DBSCAN
cluster
of solvent shell structures.
Outliers are depicted as small grey circles. The size of a point belonging to a particular cluster indicates whether it is a core point (large circle) or 
corresponds to an 
edge of the cluster (small circle). Insets depict the Fe ion (black) and surrounding oxygen atoms (hydrogen atoms are omitted for clarity), with the convex hull of the first six neighbours shown in translucent grey, for A) an octahedral configuration, B) a prism aquisition structure and C) a pyramid aquisition configuration. The first six neighbouring O atoms and the seventh closest atom are blue and red, respectively. Octahedral and non-octahedral configurations can be classified using a single $\phi^{6}_{\mathrm{norm}}$ criterion, as shown by the black dashed line separating the two clusters. The light-blue dashed line shows the value of the cutoff used in calculations of the coordination number. The data were obtained from a $150$ ns trajectory for a $[\mathrm{Fe}^{3+}]:[\mathrm{H_2O}]$ ratio of 1:597.   
(b) An idealized augmented triangular prism with vertices equidistant from the center. One of the prism apex vertex positions is coloured in red, depicting where the seventh 
molecule would be located if this were a prism aquisition structure. The other six vertices and the convex hull formed by them are grey.  
(c) 
An analogous pyramid aquisition structure. 
}
  \label{fig:figSphericity}
\end{figure*}

The normalized sphericity, $\phi^{6}_{\mathrm{norm}}$, turns out to be a robust metric for distinguishing between octahedral (OH) and non-octahedral solvation shell structures. 
Figure \ref{fig:figCoord} shows 
the time evolution of
$\phi^{6}_{\mathrm{norm}}$ as well as the coordination number
for a dynamics simulation of a solution with
$[\mathrm{Fe}^{3+}]:[\mathrm{H_2O}]$ ratio of 1:597. The normalized sphericity, $\phi^{6}_{\mathrm{norm}}$, mirrors trends in the coordination number, 
but is less sensitive to instantaneous thermal deformations. Unless explicitly stated, the simulations are carried out using the 12-6 Fe-H$_2$O parametrization with the a99SB-\emph{disp} water model \cite{Zhang2021}, with chloride counterions (see Methods section and Supplementary Section 1.1). The coordination number is calculated as the number of neighbours within a cutoff distance of $2.60$ \AA, a value selected based on the results shown in Figure \ref{fig:figSphericity}.

Figure \ref{fig:figSphericity}(a) depicts a scatter plot of $\phi^{6}_{\mathrm{norm}}$ and the distance to the seventh closest water molecule from the $\mathrm{Fe^{3+}}$ ion, $r_7$. Clusters were obtained using DBSCAN (density-based spatial clustering of applications with noise) \cite{ester1996density, Schubert2017, scikit-learn}. This is a clustering algorithm that groups data points based on their density, thereby identifying clusters of high-density regions and classifying outliers as noise. Octahedral and non-octahedral configurations form two distinct clusters, which can be differentiated simply by using a single sphericity  criterion ($\phi^{6}_{\mathrm{norm}}=0.988$), as shown by the horizontal dashed grey line in Figure \ref{fig:figSphericity}(a). The non-octahedral configurations correspond to 7-fold coordination, with the water molecules at the vertices of an augmented triangular prism. This is also referred to as the capped trigonal prism (CTP) geometry, with $C_{2v}$ symmetry \cite{Hoffmann1977, Casanova2003}. 
Such a structure
has been observed for transition metal complexes.\cite{Lewis1975, Yi2020, Haiges2007, Lin1996,Huang2010}
Our analysis shows two types of CTP configurations, which we refer to as `prism aquisition' and `pyramid aquisition' structures. They have the same shape and geometry but differ in the position of the seventh  $\mathrm{H_2O}$ molecule. 
We did not observe structures where the seventh molecule resides at one of the pyramid base vertices.   

Figure \ref{fig:figSphericity}(b)-(c) illustrates an idealized CTP structure. 
The position of the seventh furthest
molecule is highlighted in red. 
The convex hull formed by the closest six neighbours in the prism aquisition structure (Figure \ref{fig:figSphericity}(b)) tends to be less flattened, compared to that formed by the the pyramid aquisition structure (Figure \ref{fig:figSphericity}(c)). Moreover, the convex hull resembles that of a distorted octahedron in the case of the prism aquisition structure. A mechanism for the formation of the 
extended solvation shell is indicated, wherein the seventh closest water molecule tends to capture one of the prism apex vertices during its approach. A prism aquisition structure can subsequently deform or evolve into a pyramid aquisition structure. Furthermore, as the seventh furthest water molecule is ejected from the first solvation shell, it occupies a prism apex vertex. 


With the help of this $\phi^{6}_{\mathrm{norm}}$ criterion, we can differentiate between OH and CTP configurations (including the aquisition mechanism) and evaluate 
the lifetime of each state.
We posit that solvation structural properties and behaviour arise directly from two separate contributions: those from the OH state, and from the CTP state.   

The average lifetime, $\tau$, of a particular state is defined as $\tau = \frac{1}{N} \sum_{i=1}^{N}{ t_{i}^{\mathrm{end}}-t_{i}^{\mathrm{begin}} } $, for every interval $i \in [1,N] \subset \mathbb{N}$ in which the current configuration persists without switching to the other state, where $t$ represents the elapsed simulation time. This metric embodies the switching frequency between states, and is an indication of the stability of a particular state. On the other hand, the propensity of the system to exist in a particular configuration can be described by the probability, $p$, of observing that state, {\it i.e.} the ratio of the total time spent in the state, to the total simulation time. 

Both of these metrics directly influence other average structural measures, such as the coordination number (CN), the average distance of the first six neighbours from the Fe ion center ($\overline{r_{i\leq 6}}$), ion-oxygen distance (IOD), etc. In the previous literature, the CN and IOD have been reported for various potential energy functions \cite{Li2014,Li2021a,Zhang2021}. However, the simulation time has typically been short, 
on the order of $10$ ns,
which 
might not yield accurate long-time averages and adequately sample the OH and CTP states.
Consequently, we perform long simulations on the order of $100$ ns, for several ion concentrations. 

\begin{figure}
  \centering
  \includegraphics[width = \linewidth]{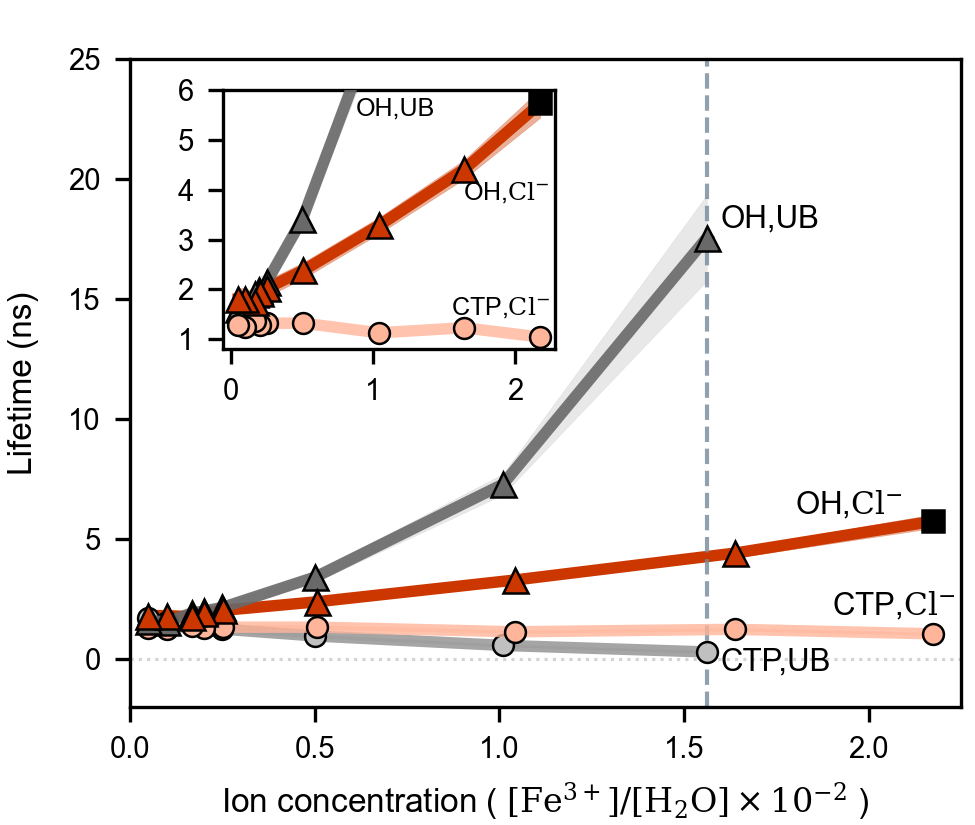}
  \caption{
Calculated lifetime, $\tau$, of the OH state and the CTP state as a function of the ion concentration, for systems with explicit counterions ($\mathrm{Cl^-}$) and systems with uniform negative background charge (UB). Data from simulations performed with explicit counterions are denoted by orange-red markers for the OH state and light orange circles for the CTP state. The OH configurations all correspond to $[\mathrm{Fe(H_2O)_6}]^{3+}$, except for the black square marker representing a concentration at which a small proportion of the \emph{trans}-$[\mathrm{FeCl_2(H_2O)_4}]^+$ was observed along with the predominant $[\mathrm{Fe(H_2O)_6}]^{3+}$. The lifetime of the OH state increases with ion concentration, while that of the CTP state decreases slightly. Therefore, the relative probability of the OH state increases with increasing ion concentration. 
 Results of simulations using uniform background charge are shown by dark grey triangles for OH and light grey for CTP. The vertical dashed grey line shows the concentration used in reported DFT based simulations.\cite{Mandal2022} 
 The use of uniform background charge gives results that are significantly different from those where explicit counterions are included,
 especially for high concentration.
The inset shows a close-up view, with a different range on the vertical axis.
}
  \label{fig:figLifetime}
\end{figure}

Our classical dynamics simulations show that the ion concentration can have a strong effect on the solvation shell. Figure \ref{fig:figLifetime} depicts changes in the average lifetime of the OH and CTP states, as the ion concentration is varied.
%
At a low concentration of $[\mathrm{Fe}^{3+}]:[\mathrm{H_2O}]$=1:597, the lifetime of the OH state, $\tau_{\mathrm{OH}}$, is $1.75$ ns. This concentration is typical for simulations based on potential energy functions 
\cite{Zhang2021, Zhang2021a}.
On the other hand, simulations based on electronic structure calculations of the energy and atomic forces, in particular DFT calculations, tend to be performed at a higher concentration \cite{Mandal2022}, for instance, a $[\mathrm{Fe}^{3+}]:[\mathrm{H_2O}]$ ratio close to of 1:61. Strikingly, $\tau_{\mathrm{OH}}$ is then $2.5$ times longer compared to the lower concentration.
However, the average lifetime of the CTP state, $\tau_{\mathrm{CTP}}$, turns out to be relatively insensitive to the ion concentration, decreases only slightly. This implies that, for a high ion concentration (obtained when small periodic simulation box size is used), the probability of observing the CTP state
is low. 

DFT calculations, furthermore, tend to employ a uniform background charge, in lieu of explicit counterions. Figure \ref{fig:figLifetime} also shows results of calculations of $\tau_{\mathrm{OH}}$ and $\tau_{\mathrm{CTP}}$ for systems with  a uniform background charge.
For low concentration, less than 1:400, the results with uniform background charge agree, within error bars, with data from simulations with explicit counterions. However, a significant difference between the two is evident at higher concentration, for example 1:64 which is typical for DFT calculations (dashed vertical line in Figure~\ref{fig:figLifetime}). We also note that, at the higher ion concentration, the value of $\tau_{\mathrm{CTP}}$ with uniform background charge 
also concomitantly decreases, thereby further reducing the probability of observing the CTP state under such conditions. Therefore, the uniform background charge tends to spuriously favour the OH state. 

These results indicate that in order to test whether the CTP state is present at the DFT level of theory, it would be necessary to carry out simulations with significantly lower ion concentration and explicit counterions instead of uniform background charge. Moreover, simulation time intervals significantly longer than a few ps would be needed. 

Our simulations also show clear evidence of ion pairing between the $\mathrm{Fe^{3+}}$ and $\mathrm{Cl^{-}}$ ions, for both the OH and CTP states; the degree of which increases as the concentration increases. 
At a concentration of 1:61, one or even two water molecules are shared between the central $\mathrm{Fe^{3+}}$ and each $\mathrm{Cl^{-}}$. Evidently, the lifetime of the OH state could, in particular, be affected by such ion pairing at high concentration. Further investigation is required to assess the effects of ion pairing. 

We also note that, for the highest concentration considered here, 
a small proportion of \textit{trans}- $\mathrm{[Fe(H}_{2}\mathrm{O})_{4}\mathrm{Cl}_{2}]^{+}$ is formed, in addition to the dominant octahedral $\mathrm{[Fe(H}_{2}\mathrm{O})_{6}]^{3+}$ configuration, and low 
occurrence
of the CTP state, $\mathrm{[Fe(H}_{2}\mathrm{O})_{7}]^{3+}$. This is in accordance with the observed trend of increased ion pairing at high concentration. The formation of the \textit{trans}-$\mathrm{[Fe(H}_{2}\mathrm{O})_{4}\mathrm{Cl}_{2}]^{+}$ structure is consistent with experimental observation of this species at a concentration 
of 1 $\mathrm{mol/dm^3}$ \cite{Persson2018}. This implies that the potential energy function employed here can
reproduce some qualitative features of the rich structural behaviour exhibited by concentrated ferric chloride solutions.  

\begin{figure}
  \centering
  \includegraphics[width = \linewidth]{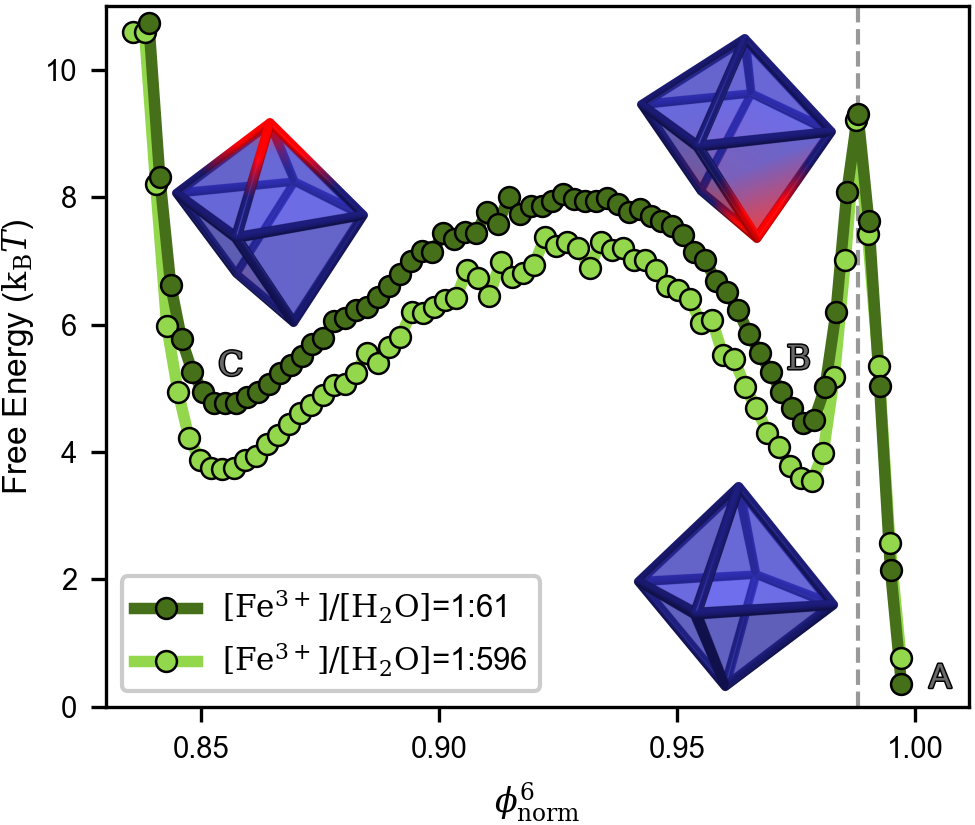}
  \caption{
Free energy as a function of the normalized sphericity, $\phi^{6}_{\mathrm{norm}}$, for 
high (dark green) and low (light green)
concentration, corresponding to a Fe$^{3+}$:H$_2$O ratio
of 1:61 and 1:596. 
Symbols have the same meaning as in Figure~\ref{eqn:sphericity}. The global minimum, represented by A, 
corresponds to
the OH state. The local minima labeled B and C depict the prism acquisition CTP state and the pyramid aquisition state. Insets show representative snapshots of each state. The dashed grey line represents the $\phi^{6}_{\mathrm{norm}}$ criterion used to differentiate between the OH and CTP states.
}
  \label{fig:fes}
\end{figure}

The free energy is calculated from the probability of the different states obtained from the classical dynamics trajectories including explicit counterions (see Supplementary). Figure~\ref{fig:fes} shows the free energy profile calculated for high 
and low concentration. 
The preferred, lowest free energy state is the OH state (A). The prism acquisition CTP state (B) and the pyramid acquisition CTP state (C) are local minima. The $\phi^{6}_{\mathrm{norm}}$ criterion used to differentiate between the CTP and OH states is depicted by a dashed grey line in Figure~\ref{fig:fes}, validating the efficacy of the criterion.

The free energy difference between the OH and CTP states is smaller for the lower concentration as compared to the higher concentration, consistent with the concomitantly lower value of $p_{\mathrm{OH}}$. We note that the free energy difference between the prism acquisition and pyramid acquisition states is small for the whole range in concentration. This is in accordance with the high frequency of switching between the CTP states, observed in the dynamics simulations. 

To further assess the stability of the CTP state, we optimized a set of eight systems comprising 64 water molecules and a Fe$^{3+}$ ion using DFT with the PBE functional approximation.
A classical dynamics simulation was first carried out for a concentration of 1:64 (dashed grey line in Figure \ref{fig:figLifetime}) 
 and eight CTP configurations were selected. 
The energy was then minimized using steepest descent to the nearest local energy minimum. The resulting configurations were subsequently used as input for DFT/PBE calculations where, again, the energy was minimized. 
%
In all cases, the energy difference between the initial and final structure in the DFT calculations is around 5~eV (i.e. 0.03~eV per atom), mainly due to the fact that bond lengths are shorter for the potential functions used here, as compared to the DFT/PBE values. 
Four of the  systems retained the CTP configuration of the solvation shell, while the rest changed into the OH state.

Additionally, a geometry optimization of an isolated $\mathrm{[Fe(H}_{2}\mathrm{O})_{7}]^{3+}$ cluster was performed
using DFT with the PBE and B3LYP functionals, starting with the CTP minimal energy configuration obtained with the 
a99SB-disp potential energy function. 
The CTP configuration of the water molecules could be maintained during careful energy minimization, 
but the energy barrier for a transformation to the OH state turns out to be low. 
Therefore, the DFT results indicate that a CTP configuration is possible but is more likely as a metastable state in `bulk' solution. 

In conclusion,
we find that a coordination number larger than 6.0, previously reported from studies using empirical potential energy functions, arises from the formation of a metastable solvation shell with 7 water molecules in a CTP geometry.  
The predominant state is, however, the commonly assumed sixfold coordinated OH configuration. 
The classical dynamics simulations 
reveal a strong dependence of the relative probability of the two states on the ion concentration. Furthermore, the use of a uniform background charge, instead of explicit counterions, tends to spuriously favour the OH state at high ion concentrations. 
The lack of CTP states in previous DFT based simulations can be attributed to the 
short timescale covered by the simulations, the high ion concentration of the simulated solutions and the use of a uniform background charge.
Our structure optimizations, performed with DFT, indicate that the CTP states could be metastable, at this level of theory as well. 
Additional experimental and theoretical studies, based on electronic structure calculations, for low ion concentrations and explicit counterions, are warranted to affirm the existence of the CTP state.
 
Further investigation is also needed to gain a better understanding of 
the extent to which rigid fixed point-charge/water models can reproduce the true interaction between the ion and the water molecules. It is reasonable to expect that a potential function including polarizability would give a more accurate description.

\section{Methods}


Classical dynamics simulations are carried out for a range of ion concentrations, using potential energy functions and the LAMMPS software\cite{Thompson2022}. The simulated system consists of Fe$^{3+}$, Cl$^{-}$ ions and H$_2$O molecules, modeled by fixed-charge rigid water models, in such a way as to maintain charge neutrality. For comparison, simulations are also carried out using uniform negative background 
instead of explicit Cl$^{-}$ ions. 
We show that averages of structural properties are equivalent for systems with the same ionic concentration, irrespective of the simulation box size (Supplementary Section 2).  

We have employed the 12-6 Lennard-Jones potential, parameterized by \citet{Zhang2021}, for modelling non-bonded non-Coulombic interactions between Fe and $\mathrm{H_2O}$, described by the a99SB-\emph{disp} water model \cite{Robustelli2018}. Parameters for chloride-chloride interactions were obtained from \citet{Smith1994}, which have previously been used in a study of $\mathrm{FeCl_{2}}$ in water \cite{Luemmen2010}. Cross-interactions involving LJ interactions were calculated using the Lorentz-Berthelot mixing rules. 
The simulations described here make use of the
Fe - a99SB-\emph{disp} water 
parameterization from \citet{Zhang2021}. Results using other 12-6 \cite{Zhang2021} and 12-6-4 \cite{Li2021a,Li2014} parameterization are provided in the Supplementary. 

About $4000$ $\mathrm{H_2O}$ molecules were used, and the ionic concentration varied by creating initial simulation configurations with different number of ions placed randomly via PACKMOL \cite{Martinez2009}. Energy minimization was first performed using $1000$ steps of steepest descent and $1000$ subsequent steps of conjugate gradient, with a timestep of $0.001$ fs. Equilibration at $300$ K was attained by performing simulations in the NPT ensemble, for at least $25$ ns. This was followed by another equilibration run in the NVT ensemble, at $300$ K, for $5.5$ ns. The production simulations were conducted in the NVT ensemble for long time intervals, ranging from $50$-$150$ ns. See Supplementary Section 1 for further simulation details.

The DFT calculations were carried out using the Perdew-Burke-Ernzerhof (PBE) \cite{PBExc} exchange correlation functional and a plane wave basis set as implemented in the VASP software \cite{
VASP_3,VASP_4}
and using the B3LYP functional\cite{Becke1988,Becke1993} and the def2-TZVP basis set\cite{B508541A} as implemented in the ORCA software.\cite{Neese2012,Neese2022} The system used in the DFT/PBE calculations consisted of 64 water molecules and an Fe$^{+3}$ ion in a cubic simulation cell of length 12.44 {\AA} with periodic boundary conditions. 
The plane-wave energy cutoff was set to 500 eV and a $\Gamma$-point only sampling of the first Brillouin zone was used. The energy minimization of atomic configurations was carried out until the magnitude of atomic forces had dropped below 0.02~eV/{\AA}.

Snapshots and visuals of solvation systems were created using \texttt{solvis} (\url{https://github.com/amritagos/solvis}). Scripts for calculating the sphericity of solvation shells are also provided in the same. 

\begin{acknowledgement}
This work was funded by the Icelandic Research Fund (grants 228615-051 and 207283-053).
The calculations were performed using compute resources provided by the Icelandic Research Electronic Infrastructure (IREI).
We are grateful to Moritz Sallermann, Rohit Goswami and Kathleen A. Schwarz for fruitful discussions.
\end{acknowledgement}

\begin{suppinfo}

The authors confirm that the data supporting the findings of this study are available within the article and/or its supplementary materials.
The supporting information contains a description of the simulation methodology for molecular dynamics (MD) simulations with empirical potential functions as well as DFT simulation details, data on the
validation of ion concentration effects as well as finite size effects; comparison of the performance of various classical non-bonded potentials; details on free energy calculations performed.
Minimized configurations of the CTP state for isolated ion-water clusters are available at \url{https://zenodo.org/doi/10.5281/zenodo.10680196}.

\end{suppinfo}

\providecommand{\latin}[1]{#1}
\makeatletter
\providecommand{\doi}
  {\begingroup\let\do\@makeother\dospecials
  \catcode`\{=1 \catcode`\}=2 \doi@aux}
\providecommand{\doi@aux}[1]{\endgroup\texttt{#1}}
\makeatother
\providecommand*\mcitethebibliography{\thebibliography}
\csname @ifundefined\endcsname{endmcitethebibliography}
  {\let\endmcitethebibliography\endthebibliography}{}

\afterpage{\blankpage}



\section*{Supplementary material}

Table S1: Comparison of the average octahedral lifetime, $\tau_{\text{OH}}$, for systems with different sizes, for various ion concentrations, expressed as $[\text{Fe}^{3+}]/[\text{H}_2\text{O}]$ ratios. The so-called small systems consist of a single Fe ion, along with water molecules and counterions, corresponding to the prevailing ion concentration. The large system contains about $4000$ to $5000$ water molecules in each case, with the same $[\text{Fe}^{3+}]/[\text{H}_2\text{O}]$ ratio as in the smaller system. The standard errors are provided in brackets. The 12-6 parameterization with the a99SB-*disp* water model[4] was used for these simulations.

| $[\text{Fe}^{3+}]/[\text{H}_2\text{O}]$ | $\tau_{\text{OH}}$ (ns) | |
| --- | --- | --- |
| | Small system | Large system |
| 1:996 | 1.701(0.229) | 1.801(0.225) |
| 1:596 | 1.867(0.300) | 1.757(0.181) |
| 1:396 | 1.611(0.274) | 2.023(0.166) |
| 1:61 | 3.825(0.420) | 4.41(0.192) |

system size is irrelevant insofar as the average solvation structure is concerned.

## 3 Comparison of Various Classical Non-Bonded Potentials

The stability of a state is reflected by the value of its lifetime, $\tau$. On the other hand, the propensity of the system to exist in a particular configuration can be described by the probability of observing that state. We define the 'octahedral probability', or $p_{\text{OH}}$, as the ratio of the total time spent in the octahedral (OH) state, to the total simulation time of the system. This roughly translates to the probability of observing the OH state at any given point in time. Since this is a two-state system, the probability of the CTP (capped trigonal prism) state is given by $p_{\text{CTP}} = 1 - p_{\text{OH}}$.

Despite the fact that the various potential energy functions, considered in this work (described in Section 1.1), all perform similarly well, with respect to the hydration free energy and ion-oxygen distance, the relative stability and frequency of switching between the OH and CTP states vary significantly.

Table S2 contains values of $p_{\text{OH}}$ and $\tau_{\text{OH}}$, at a low and high ion concentration, for the

Table S2: Values of the effective probability, $p_{\text{OH}}$ (%), and the average lifetime, $\tau_{\text{OH}}$ (ns), of the OH state, at low and high ion concentrations (%), for the Fe-water parameterizations for the 12-6 potential[4] with the a99SB-*disp* water model and the OPC model, and the 12-6-4 potential[5,6] with the OPC, OPC3 and TIP4P-Ew water models. The low concentration and high concentration correspond to ion concentrations with $[\text{Fe}^{3+}]/[\text{H}_2\text{O}]$ ratios of 1:1996 and 1:61, respectively.

| Model | Low concentration (1:1996) | | High concentration (1:61) | |
|---|---|---|---|---|
| | $p_{\text{OH}}$ (%) | $\tau_{\text{OH}}$ (ns) | $p_{\text{OH}}$ (%) | $\tau_{\text{OH}}$ (ns) |
| 12-6 a99SB-*disp*[4] | 57 | 1.803(0.225) | 80 | 4.411(0.192) |
| 12-6 OPC[4] | 34 | 0.809(0.109) | 42 | 1.237(0.185) |
| 12-6-4 OPC[5] | 32 | 0.200(0.004) | 36 | 0.213(0.019) |
| 12-6-4 OPC3[5] | 19 | 0.109(0.006) | 25 | 0.161(0.047) |
| 12-6-4 TIP4P-Ew[6] | 7 | 0.067(0.006) | 11 | 0.123(0.057) |

various classical non-bonded potentials.

Figure S1 depicts the energy barrier for the isolated ion-water clusters, determined using climbing image nudged-elastic band (CI-NEB)[25–27] calculations for the 12-6 potential with the a99SB-*disp* water model, and the 12-6-4 potential with the OPC, OPC3 and TIP4P-Ew models, respectively.

The 12-6-4 potentials[5,6] tend to produce both octahedral (OH) and capped trigonal prism (CTP) states with lower lifetimes compared to the 12-6 potentials: in other words, the system rattles between the states more rapidly for the 12-6-4 potentials (particularly apparent when comparing the 12-6-4 OPC[5] and 12-6 OPC parameterizations[4]). This can be qualitatively appraised by examining the shape of the energy barriers, which are flatter compared to the higher and steeper barriers for the 12-6 potentials (Figure S1). However, $p_{\text{OH}}$ values are more-or-less in agreement for both the 12-6 OPC parameterization and the 12-6-4 OPC parameterization (Table S2). Therefore, we expect the relative distribution of configurations throughout the trajectory, in both the OH and CTP states, to be similar for both classes of potentials.

The data from Table S2 reveal that the probability of observing the OH state is greater for higher ion concentrations, for both the 12-6 and 12-6-4 potential parameterizations,

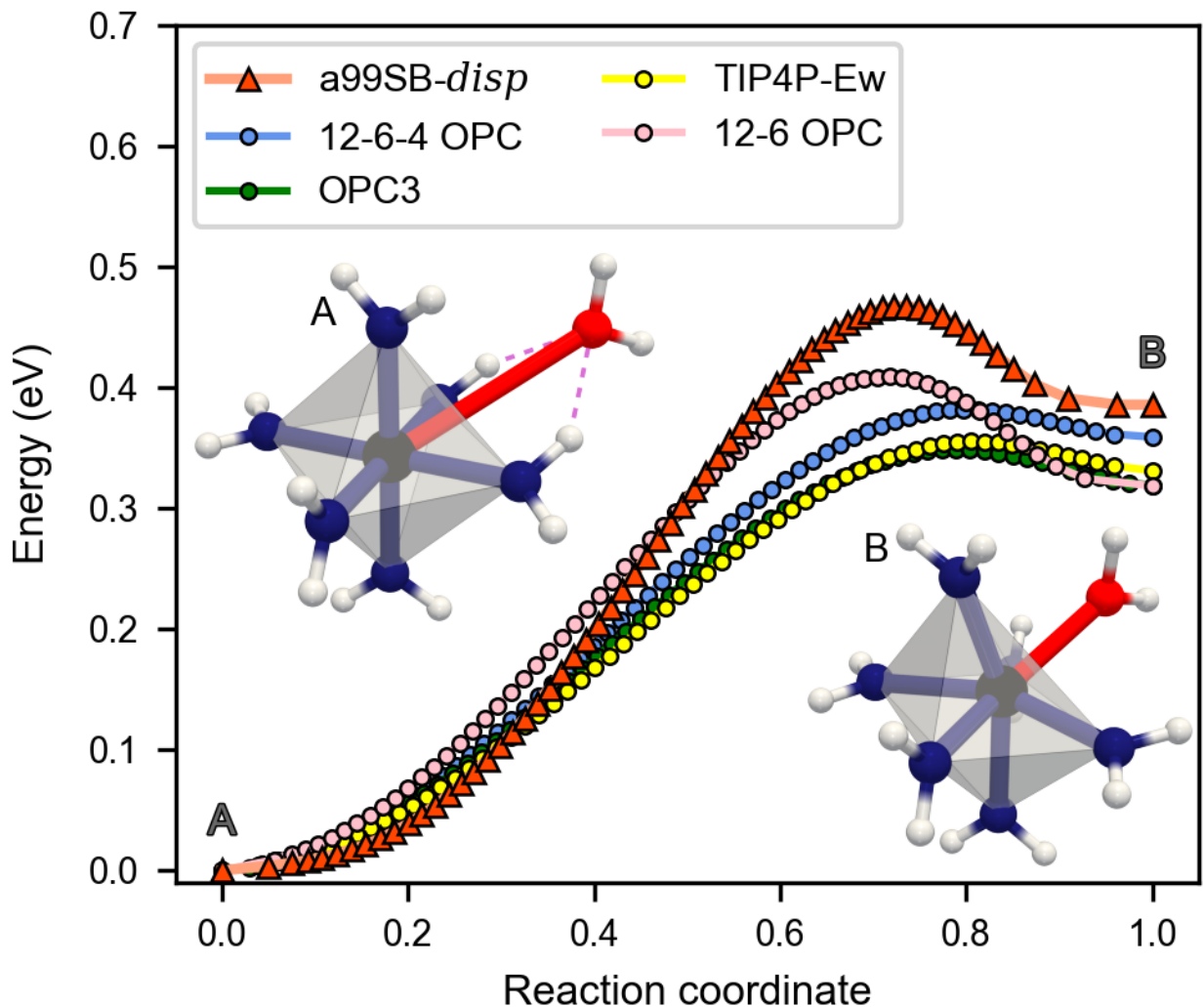

Figure S1: Energy profile along the minimum energy path obtained from CI-NEB calculations for the transition between the OH state (A) and the CTP state (B), using various potential energy functions: 12-6 potential with the a99SB-*disp* water model[4] and OPC models,[4] respectively, and the 12-6-4 potential with the OPC,[5] OPC3[5] and TIP4P-Ew[6] water models. An energy barrier exists between the local minima corresponding to the two states. Each system consists of a single $Fe^{3+}$ ion and seven $H_2O$ molecules. The central $Fe^{3+}$ ion, H atoms, the six closest O atoms and the seventh O atom are coloured in black, light grey, blue and red, respectively. Hydrogen bonds formed by the seventh water molecule are depicted as pink dashed lines.

although the difference is smaller for the 12-6-4 models. In the case of the 12-6 potential parameterization with the a99SB-*disp* water model, the lifetimes strongly reflect an increase with increase in ion concentration.

**Table S3:** Comparison of various structural parameters for energy minimized configurations of isolated ion-water clusters containing seven water molecules, modelled using the $Fe^{3+}$-$H_2O$ parameterizations for the 12-6 potential with the a99SB-*disp*[4] and OPC[4] water models, respectively, and the 12-6-4 potential with the OPC,[5] OPC3[5] and TIP4P-Ew[6] water models, as well as using DFT calculations with the PBE and B3LYP functional approximations. The average of the distance (in Å) from the central ion to the six closest neighbours, $\overline{r_{i\leq 6}}$, and the distance of the ion from the seventh water molecule, $r_7$, are listed.

| Model | OH config. | | CTP config. | |
|---|---|---|---|---|
| | $\overline{r_{i\leq 6}}$ | $r_7$ | $\overline{r_{i\leq 6}}$ | $r_7$ |
| 12-6 a99SB-*disp*[4] | 2.026 | 3.606 | 2.091 | 2.162 |
| 12-6 OPC[4] | 2.019 | 3.537 | 2.088 | 2.153 |
| 12-6-4 OPC[5] | 2.020 | 3.508 | 2.091 | 2.208 |
| 12-6-4 OPC3[5] | 1.999 | 3.468 | 2.067 | 2.182 |
| 12-6-4 TIP4P-Ew[6] | 1.980 | 3.454 | 2.049 | 2.174 |
| DFT/PBE | 2.055 | 3.620 | 2.121 | 2.253 |
| DFT/B3LYP | 2.055 | 3.767 | 2.124 | 2.278 |

A very striking point to note here, regarding the performance of both 12-6 and 12-6-4 parameterizations with the OPC, OPC3 and TIP4P-Ew water models, is that the system spends significantly less than 50% of the total simulation time in the OH state, at all ion concentrations investigated. This behaviour is egregiously spurious, even if we consider the existence of the CTP configuration as a rare state. We can link the relative performance of the parameterizations to the faithfulness of the $\overline{r_{i\leq 6}}$ values in reproducing of the experimental ion-oxygen distance, exhibited by isolated ion-water clusters in the OH state (Table S3). A causal link to qualities of the specific water model employed is indicated, but a more thorough investigation is beyond the scope of this work.

It is clear, both from the greater $p_{\text{OH}}$ and $\tau_{\text{OH}}$ values, that the Fe-water parameterization with the a99SB-*disp* water model exhibits OH configurations that persist longer than those for the other models, and that the probability of the OH state is greater. In other words, the OH configuration appears to be more stable and is the preferred state for the a99SB-*disp*

water parameterization, for all ion concentrations considered here (but especially at higher concentrations). This more closely approximates expected behaviour.

## 4  Free Energy Calculations

We have shown how the normalized sphericity, $\phi^6_{\text{norm}}$, is a robust criterion for differentiating between OH and CTP states. We consider $\phi^6_{\text{norm}}$ to be our collective variable (CV), represented by $s$, noting that $s$ is a function of system coordinates. We have used the canonical ensemble for our production runs.

Therefore, the Helmholtz free energy, $F(s)$, is given by:[28]

$$F(s) = -k_{\text{B}} T \ln P(s), \tag{S1}$$

where $P(s)$ is the equilibrium probability density for $s$, $k_{\text{B}}$ is the Boltzmann constant, and $T$ is the temperature.



\end{document}